\documentclass[10pt]{article}

\usepackage{amsmath, bm, mathrsfs, amssymb}
\usepackage[dvipsnames]{xcolor}
\usepackage[hidelinks,
            colorlinks=true,
            linkcolor=NavyBlue,
            citecolor=darkgray,
            urlcolor=NavyBlue]{hyperref}
\usepackage{pifont}
\usepackage{graphicx}
\usepackage{doi}

\usepackage{natbib}
\setcitestyle{round,citesep={,},aysep={}}

\newcommand{\documentname}{\textsl{Note}}
\newcommand{\sectionname}{Section}
\newcommand{\equationname}{equation}
\newcommand{\notename}{Note}

\newcommand{\acronym}[1]{{\small{#1}}}
\newcommand{\code}[1]{\texttt{#1}}
\newcommand{\foreign}[1]{\textsl{#1}}

\newcommand{\hquad}{~~}
\newcommand{\given}{\,|\,}
\newcommand{\dd}{\mathrm{d}}
\newcommand{\T}{^{\!\mathsf{T}\!}}
\newcommand{\inv}{^{-1}}

\renewcommand{\vector}[1]{\boldsymbol{#1}}
\newcommand{\tensor}[1]{\mathbf{#1}}
\renewcommand{\matrix}[1]{\mathsf{#1}}
\newcommand{\normal}{\mathcal{N}\!\,}

\newcommand{\va}{\vector{a}}
\newcommand{\vb}{\vector{b}}
\newcommand{\vm}{\vector{m}}
\newcommand{\vx}{\vector{x}}
\newcommand{\vy}{\vector{y}}

\newcommand{\vmu}{\vector{\mu}}
\newcommand{\veta}{\vector{\eta}}
\newcommand{\vtheta}{\vector{\theta}}
\newcommand{\tA}{\tensor{A}}
\newcommand{\tB}{\tensor{B}}
\newcommand{\tC}{\tensor{C}}

\newcommand{\tI}{\tensor{I}}
\newcommand{\tQ}{\tensor{Q}}
\newcommand{\tS}{\tensor{S}}
\newcommand{\tH}{\tensor{H}}
\newcommand{\tV}{\tensor{V}}
\newcommand{\tLambda}{\tensor{\Lambda}}
\newcommand{\mM}{\matrix{M}}

\newcommand{\mU}{\matrix{U}}
\newcommand{\mV}{\matrix{V}}
\newcommand{\bP}{\ensuremath{\textrm{\ding{80}}}} 
\newcommand{\bH}{\ensuremath{\textrm{\ding{114}}}} 

\sloppypar

\usepackage{marginfix} 
\newcounter{marginnote}
\renewcommand{\footnote}[1]{\refstepcounter{marginnote}\textsuperscript{\themarginnote}\marginpar{\color{darkgray}\raggedright\footnotesize\textsuperscript{\themarginnote}#1}}
\newcommand{\tfigurerule}{\rule{0pt}{1ex}\\ \rule{\marginparwidth}{0.5pt}\\ \rule{0pt}{0.25ex}}
\newcommand{\bfigurerule}{\rule{0pt}{0.25ex}\\ \rule{\marginparwidth}{0.5pt}\\ \rule{0pt}{1ex}}
\renewcommand{\caption}[1]{\parbox{\marginparwidth}{\footnotesize\refstepcounter{figure}\textbf{\figurename~\thefigure}: {#1}}}

\begin{document}

\section*{Data Analysis Recipes:\\
  Products of multivariate Gaussians\\
  in Bayesian inferences}

\noindent\textbf{David W. Hogg}\footnote{%
  The authors would like to thank
  Will Farr (Stony Brook),
  Dan Foreman-Mackey (Flatiron),
  Rodrigo Luger (Flatiron),
  Hans-Walter Rix (MPIA),
  and
  Sam Roweis (deceased),
  for help with all these concepts.
  This project was developed
  in part at \textsl{AstroHackWeek 2016}, which was
  hosted by the Moore--Sloan Data Science Environment.
  This research was supported by the National Science Foundation and National Aeronautics and Space Administration.
  The source text, example IPython notebooks, and data files used below are
  available via \href{https://doi.org/10.5281/zenodo.3855689}{Zenodo archive}
  \citep{zenodo}.

}\\
{\footnotesize%
  \textsl{Center for Cosmology and Particle Physics, Dept.\ Physics, New York University}\\
  \textsl{Max-Planck-Institut f\"ur Astronomie, Heidelberg}\\
  \textsl{Flatiron Institute, a division of the Simons Foundation}%
}

\medskip\noindent\textbf{Adrian~M.~Price-Whelan}\\
{\footnotesize%
  \textsl{Flatiron Institute, a division of the Simons Foundation}%
}

\medskip\noindent\textbf{Boris Leistedt}\\
{\footnotesize%
  \textsl{Department of Physics, Imperial College, London}\\
  \textsl{Center for Cosmology and Particle Physics, Dept.\ Physics, New York University}%
}

\paragraph{Abstract:}
A product of two Gaussians---or normal distributions---is another Gaussian.
That's a valuable and useful fact!
Here we use it to derive a refactoring of a common product of
multivariate Gaussians:
The product of a Gaussian likelihood times a Gaussian prior, where some or all
of those parameters enter the likelihood only in the mean and only linearly.
That is, a linear, Gaussian, Bayesian model.
This product of a likelihood times a prior pdf can be refactored into a product of a
marginalized likelihood (or a Bayesian evidence) times a posterior pdf, where
(in this case) both of these are also Gaussian.
The means and variance tensors of the refactored Gaussians are straightforward
to obtain as closed-form expressions;
here we deliver these expressions, with discussion.
The closed-form expressions can be used to speed up and improve the precision
of inferences that contain linear parameters with Gaussian priors.
We connect these methods to inferences that arise frequently in physics
and astronomy.

If all you want is the answer, the question is posed and answered at the
beginning of \sectionname~\ref{sec:problemsolution}.
We show two toy examples, in the form of worked exercises, in
\sectionname~\ref{sec:examples}.
The solutions, discussion, and exercises in this \documentname\ are aimed at
someone who is already familiar with the basic ideas of
Bayesian inference and probability.

\section{Inferences with linear parameters}

It is common in physics, astronomy, engineering, machine learning, and many other fields that likelihood functions
(probabilities of data given parameters) are chosen to be Gaussian
(or normal\footnote{In this \documentname, we obey physics and astronomy
conventions and refer to the normal pdf as the Gaussian pdf. We realize that it
is irresponsible to name things after people when those same things also have generic,
descriptive names. On the other hand, ``normal'' isn't the finest name either.
Perhaps Gaussian pdfs should be called ``central'' since they are produced by
the central limit theorem. Anyway, we will continue with the name ``Gaussian''
despite our own reservations. We apologize to our reader.}):
One reason is that a likelihood function is basically a noise model,
and it is often case that the noise is treated as Gaussian.
This assumption for the likelihood function is \emph{accurate} when the noise
model has benefitted from the central limit theorem.
This is true, for example, when the noise is thermal, or when the
noise is shot noise and the numbers (numbers of detected photons or other
particles) are large.
Another reason that the likelihood function is often treated as
Gaussian is that Gaussians are generally \emph{tractable}:
Many computations we like to perform on Gaussians, like integrals and
derivatives and optimizations, have closed-form solutions.
Even when we don't use the closed-form solutions, there are many
contexts in which Gaussians lead to convex optimizations,
providing guarantees to resulting inferences.

It is also common in physics and astronomy that models for data
include parameters such that the expectation value for the data (in,
say, a set of repeated experiments) is linearly proportional to some
subset of the parameters.
This is true, for example, when we fit a histogram of \textsl{Large Hadron
  Collider} events affected by the Higgs boson,\footnote{See, for example,
  \cite{atlas}, and \cite{cms}, and references therein.}
where the expected number of counts in each
energy bin is proportional to a linear combination of the amplitudes
of various backgrounds and some coupling to the Higgs.
Another linear-parameter context, for example, arises when we fit for the radial-velocity
variation of a star in response to a faint, orbiting companion.\footnote{See,
  for an example in our own work, \cite{Price-Whelan:2017, Price-Whelan:2020}.
  That project and those papers would have been
  impossible without the speed-ups provided by the expressions derived in this
  \documentname. Indeed, the writing of this \documentname\ was motivated by
  the work presented in those papers. (Okay full disclosure: It was motivated in
  part by a \emph{mistake} made by DWH in one of those papers!)}
In this problem, the expectation of the
radial-velocity measurements depends linearly on the binary system
velocity and some combination of masses and system inclination (with
respect to the line of sight).
In both of these cases, there are both linear parameters (like the
amplitudes) and nonlinear parameters (like the mass of the Higgs, or
the orbital period of the binary-star system).
In what follows, we will spend our energies on the linear parameters,
though our work on them is in service of learning the nonlinear
parameters too, of course.

In Bayesian inference contexts, the ``models'' to which we are referring are
expressions for likelihood functions and prior pdfs; these are the things
that will be Gaussians here.
Bayes' theorem is often written as a ratio of probability density functions
(pdfs in what follows), but it can also be written as a pdf factorization:\footnote{For
  a tutorial on probability factorizations, see \cite{probcalc}.}
\begin{equation}\label{eq:bayes}
p(\vy,\vtheta\given\bH) = p(\vy\given\vtheta,\bH)\,p(\vtheta\given\bH) = p(\vtheta\given\vy,\bH)\,p(\vy\given\bH)
\end{equation}
where
$p(\vy,\vtheta\given\bH)$ is the joint probability of data $\vy$ and
parameters $\vtheta$ given your model assumptions and hyper parameters
(symbolized jointly as $\bH$),\footnote{%
  In this \documentname, we typeset different mathematical objects according to their mathematical or transformation properties.
  We typeset vectors (which are column vectors) as $\va, \vb, \vtheta$,
  we typeset variance tensors (which in this case are square, non-negative semi-definite matrices) as $\tC, \tLambda$,
  we typeset other matrices (which will in general be non-square) as $\mM, \mU$,
  and we typeset blobs or unstructured
  collections of information as $\bH, \bP$.
  Related to this typography is an implicit terminology:
  We distinguish variance tensors from matrices.
  This distinction is somewhat arbitrary, but the strong constraints on the
  variance tensors (non-negative, real eigenvalues) make them special beasts,
  with special geometric properties, like that they can be used as metrics
  in their respective vector spaces.}
$p(\vy\given\vtheta,\bH)$ is the likelihood, or probability of data $\vy$
given parameters (and assumptions),
$p(\vtheta\given\bH)$ is the prior pdf for the parameters $\vtheta$,
$p(\vtheta\given\vy,\bH)$ is the posterior pdf for the parameters $\vtheta$
given the data,
and
$p(\vy\given\bH)$ is the pdf for the data, marginalizing out all of the linear
parameters (hereafter, we refer to this as the \textsl{marginalized
likelihood}\footnote{In the case that the problem has no parameters other than
the linear parameters $\vtheta$, this term, $p(\vy \given \bH)$, is sometimes
called the \textsl{Bayesian evidence} or the \textsl{fully marginalized
likelihood}.}).

If the likelihood is Gaussian, and the expectation of the data depends linearly
on the parameters, and if we choose the prior pdf to also be Gaussian, then
all the other pdfs (the joint, the posterior, and the marginalized likelihood)
all become Gaussian too.
The main point of this \documentname\ is that the means and variances of these
five Gaussians are all related by simple, closed-form expressions, given below.
One consequence of this math is that \emph{if} you have a Gaussian
likelihood function, and \emph{if} you have a subset of parameters that are
linearly related to the expectation of the data, \emph{then} you can obtain both
the posterior pdf $p(\vtheta\given\vy,\bH)$ and the marginalized likelihood
$p(\vy\given\bH)$ with closed-form transformations of the means and variances of
the likelihood and prior pdf.

A currently popular data-analysis context in which Gaussian likelihoods are
multiplied by Gaussian priors is \textsl{Gaussian processes} (\acronym{GP}s),
which is a kind of
non-parametric fitting in which a kernel function sets the flexibility of a
data-driven model. A full discussion of \acronym{GP}s is beyond the scope of this
\documentname, but excellent discussions abound.\footnote{We like the free
  book by \cite{Rasmussen:2005}.}
The math below can be applied in many \acronym{GP} contexts.
Indeed, most linear model fits of the kind we describe below can be translated
into the language of \acronym{GP}s, because any noise process that delivers
both a prior pdf and a likelihood with Gaussian form is technically identical
to a (probably non-stationary) \acronym{GP}.
We leave that translation as an exercise to the ambitious
reader.\footnote{If you want a cheat sheet,
  we come close to performing this translation in \cite{luger}.}

\section{Marginalization by refactorization}

Imagine that we are doing an inference using data $\vy$ (which is a
$N$-dimensional vector, say).
We are trying to learn linear parameters $\vtheta$ (a $K$-dimensional vector)
and also nonlinear parameters $\bP$ (an arbitrary vector, list, or
blob).\footnote{Here, $\bP$ represents the nonlinear parameters \emph{and}
  assumptions or hyper parameters. That is, it contains everything on which
  the linear model is conditioned, including not just nonlinear parameters
  but also investigator choices. Note our subjectivism here!}
Whether we are Bayesian or frequentist, the inference is based on
a likelihood function, or probability for the data given parameters
\begin{equation}
\mbox{\small likelihood:} \hquad p(\vy\given\vtheta,\bP) \hquad.
\end{equation}

Now let's imagine that the parameters $\vtheta$ are either nuisance
parameters, or else easily marginalized, so we want to marginalize
them out.
This will leave us with a lower-dimensional marginalized likelihood
function
\begin{equation}
\mbox{\small marginalized likelihood:} \hquad p(\vy\given\bP) \hquad.
\end{equation}
That's good, but the marginalization comes at a cost:
We have to become Bayesian, and we have to choose a prior
\begin{equation}
\mbox{\small prior on nuisance parameters:} \hquad p(\vtheta\given\bP) \hquad.
\end{equation}
This is the basis for the claim\footnote{A claim that perhaps hasn't been made clearly
  yet, but will eventually be by at least one of these authors.} that Bayesian
inference requires a likelihood function, and priors on the nuisance parameters.
It does not require a prior on everything, contrary to some statements
in the literature.\footnote{It is very common for papers or projects with
  Bayesian approaches to claim that the
  goal of Bayesian inference is to create posterior pdfs. That isn't correct.
  Different Bayesian inferences have different objectives. The fundamental
  point
  of Bayesian inference is that consistently held beliefs obey the rules of
  probability. That, in turn, says that if you want
  to communicate to \emph{others} things useful to the
  updating of \emph{their} beliefs, you want to communicate about your likelihood.
  Your posterior pdf isn't all that useful to them!\label{note:lf}}
We have said ``$p(\vtheta\given\bP)$'' because this prior pdf may depend on
the nonlinear parameters $\bP$, but it certainly doesn't have to.
Armed with the likelihood and prior---if you want it---you can construct
the posterior pdf for the linear parameters
\begin{equation}
\mbox{\small posterior for nuisance parameters:} \hquad p(\vtheta\given\vy,\bP) \hquad.
\end{equation}

To perform a marginalization of the likelihood, we have two choices.
We can either do an integral:
\begin{equation}\label{eq:integral}
p(\vy\given\bP) = \int p(\vy\given\vtheta,\bP)\,p(\vtheta\given\bP)\,\dd\vtheta
\hquad,
\end{equation}
where the integral is implicitly over the entire domain of the
linear parameters $\vtheta$ (or the entire support of the prior).
Or we can re-factorize the expression using Bayes' theorem:
\begin{equation}
p(\vy\given\vtheta,\bP)\,p(\vtheta\given\bP)
 = p(\vtheta\given\vy,\bP)\,p(\vy\given\bP)
\hquad.
\end{equation}
That is, in certain magical circumstances it is possible to do this
re-factorization without explicitly doing any integral.
When this is true, the marginalization is sometimes far easier than
the relevant integral.

The point of this \documentname\ is that this magical circumstance
arises when the two probability
distributions---the likelihood and the prior---are both Gaussian in
form, and when the model is linear over the parameters we would like to marginalize over.
In detail we will assume
\begin{enumerate}
\item
the likelihood $p(\vy\given\vtheta,\bP)$ is a Gaussian in $\vy$,
\item
the prior $p(\vtheta\given\bP)$ is a Gaussian in $\vtheta$,
\item
the mean of the likelihood Gaussian depends linearly on the linear
parameters $\vtheta$, and
\item
the linear parameters $\vtheta$ don't enter the likelihood anywhere
other than in the mean.
\end{enumerate}
In equations, this becomes:
\begin{equation}
p(\vy\given\vtheta,\bP) = \normal(\vy\given\mM\cdot\vtheta,\tC)
\end{equation}
\begin{equation}
p(\vtheta\given\bP) = \normal(\vtheta\given\vmu,\tLambda)
\end{equation}
\begin{equation}\label{eq:Gaussian}
\normal(\vx\given\vm,\tV) \equiv \frac{1}{||2\pi\,\tV||^{1/2}}\,\exp\left(-\frac{1}{2}\,[\vx-\vm]\T \cdot \tV\inv \cdot [\vx - \vm]\right)
\hquad,
\end{equation}
where $\normal(\vx\given\vm,\tV)$ is the multivariate Gaussian
pdf\footnote{Check out what we did with the ``$2\pi$'' in the determinant in
  \equationname~(\ref{eq:Gaussian}): We wrote an expression for the
  multivariate Gaussian that never makes any reference to the
  dimension $d$ of the $\vx$-space. Most expressions in the literature
  have a pesky $d/2$ in them, which is ugly and implies some need for
  code or equations to know the dimension explicitly, even though all
  the terms (determinant, inner product) are coordinate-free scalar
  forms. If you use the expression as we have written it here, you never
  have to explicitly access the dimensions.} for a vector $\vx$
given a mean vector $\vm$ and a variance tensor $\tV$,
$\mM$ is a $N\times K$ rectangular design matrix (which depends, in
general, on the nonlinear parameters $\bP$),
$\tC$ is a $N\times N$ covariance matrix of uncertainties for the
data (diagonal if the data dimensions are independent).
That is, the likelihood is a Gaussian with a mean that depends
linearly on the parameters $\vtheta$, and
$\vmu$ and $\tLambda$ are the $K$-vector mean and $K\times K$ variance tensor
for the Gaussian prior.

\marginpar{\tfigurerule\\
  \includegraphics[width=\marginparwidth, trim=0ex 0ex 0ex 0.5in, clip]{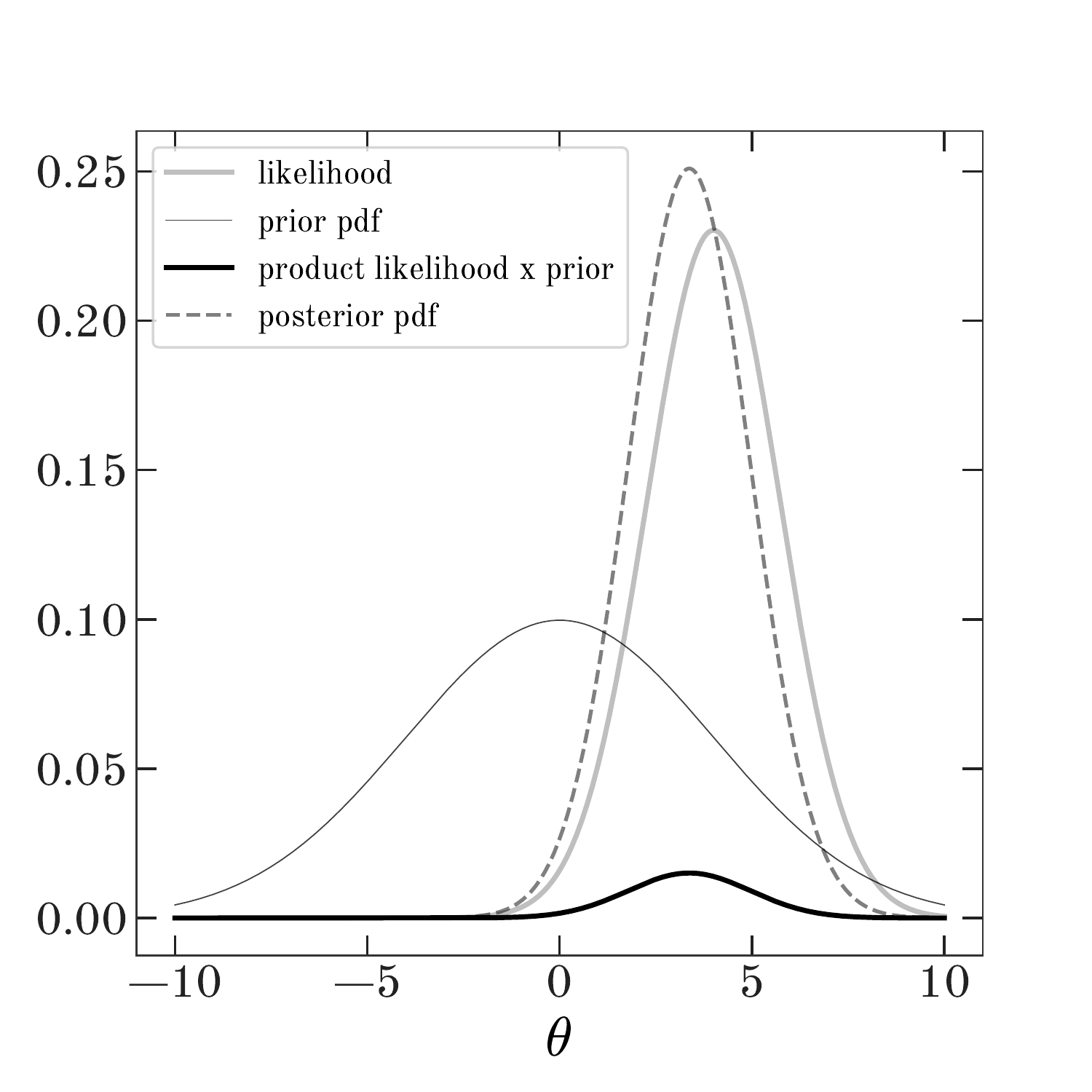}
  \caption{The one-dimensional case: If the prior pdf and the likelihood are both
    Gaussian in a single parameter, their product (and hence the posterior pdf) is
    also Gaussian, with a narrower width (smaller variance) than
    either the prior pdf or the likelihood.\label{fig:oned}}\\
  \bfigurerule}
In this incredibly restrictive---but also surprisingly
common---situation, the re-factored pdfs $p(\vtheta\given\vy,\bP)$
(the posterior for the linear parameters, conditioned on
the nonlinear parameters in $\bP$) and $p(\vy\given\bP)$ (the
marginalized likelihood, similarly conditioned) will also both be Gaussian.
We will solve this problem for general multivariate Gaussians in spaces
of different dimensionality (and units) but the one-dimensional case is
illustrated in \figurename~\ref{fig:oned}.
Obtaining the specific form for the general Gaussian product is the object of this
\documentname.

\section{Products of two Gaussians}\label{sec:problemsolution}

On the internets, there are many documents, slide decks, and videos
that explain products of Gaussians in terms of other Gaussians.\footnote{Two
  good examples are \cite{roweis}, and \cite{cookbook}. The closest---that we know
  of---to a
  discussion with the generality of
  what is shown here is perhaps our own previous contribution \cite{luger}.}
The vast majority of these consider either the univariate case (where
the data $\vy$ and the parameter $\vtheta$ are both simple scalars, which
is not useful for our science cases), or the same-dimension case (where the data
$\vy$ and the parameter vector $\vtheta$ are the same length, which never
occurs in our applications).
Here we solve this problem in the general case:\footnote{We solve this general
  case, but we are not claiming \emph{priority} in any sense: This mathematics
  has been understood for many many decades or even centuries. This \documentname\ is
  a pedagogical contribution, not a research contribution.}
The inputs are multivariate (vectors) and the two Gaussians we are
multiplying live in spaces of different dimensions.
That is, we solve the following problem:

\paragraph{Problem:}
Find $K$-vector $\va$, $K\times K$ variance tensor $\tA$, $N$-vector $\vb$,
and $N\times N$ variance tensor $\tB$ such that
\begin{equation}\label{eq:problem}
\normal(\vy\given\mM\cdot\vtheta,\tC)\,\normal(\vtheta\given\vmu,\tLambda)
 = \normal(\vtheta\given\va,\tA)\,\normal(\vy\given\vb,\tB) \hquad,
\end{equation}
and such that $\va$, $\tA$, $\vb$, and $\tB$ don't depend on $\vtheta$ at all.
Note that
$\vy$ is a $N$-vector,
$\mM$ is a $N\times K$ matrix,
$\vtheta$ is a $K$-vector,
$\tC$ is a $N\times N$ non-negative semi-definite variance tensor,
$\vmu$ is a $K$-vector,
and
$\tLambda$ is a $K\times K$ non-negative semi-definite variance tensor.

\paragraph{Solution:}
\begin{equation}\label{eq:A}
\tA\inv = \tLambda\inv + \mM\T \cdot \tC\inv \cdot \mM
\end{equation}
\begin{equation}\label{eq:a}
\va = \tA \cdot (\tLambda\inv \cdot \vmu + \mM\T \cdot \tC\inv \cdot \vy)
\end{equation}
\begin{equation}\label{eq:B}
\tB = \tC + \mM \cdot \tLambda \cdot \mM\T
\end{equation}
\begin{equation}\label{eq:b}
\vb = \mM \cdot \vmu
\hquad.
\end{equation}
This is the complete solution to the problem, and constitutes the main point
of this \documentname.
For completeness, we will give some discussion!

\paragraph{Proof:}
The two sides of \equationname~(\ref{eq:problem}) are identical if two things
hold.
The first thing is that the determinant products must be equal:
\begin{equation}
||\tC||\,||\tLambda|| = ||\tA||\,||\tB||
\hquad,
\end{equation}
because the determinants are involved in the normalizations of the
functions.
This equality of determinant products follows straightforwardly from
the matrix determinant lemma\footnote{See, for example, \cite{Wiki:MDL}, and \cite{Harville:2011}.}
\begin{equation}\label{eq:detlemma}
||\tQ + \mU\cdot\mV\T|| = ||\tI + \mV\T\cdot\tQ\inv\cdot\mU||\,||\tQ||
\hquad,
\end{equation}
where $\mU$ and $\mV$ can be rectangular, and $\tI$ is the correct-sized identity matrix.
This identity implies that
\begin{equation}
||\tA\inv|| = ||\tI + \mM\T\cdot\tC\inv\cdot\mM\cdot\tLambda||\,||\tLambda\inv||
\end{equation}
\begin{equation}
||\tB||     = ||\tI + \mM\T\cdot\tC\inv\cdot\mM\cdot\tLambda||\,||\tC||
\hquad,
\end{equation}
where we had to apply the identity twice to get the $||\tA\inv||$ expression.
We can ratio these as follows to prove this first thing:
\begin{equation}
||\tA||\,||\tB||
 = \frac{||\tB||}{||\tA\inv||}
 = \frac{||\tC||}{||\tLambda\inv||}
 = ||\tC||\,||\tLambda||
\hquad.
\end{equation}

The second thing required for the proof is that the quadratic scalar form
\begin{equation}\label{eq:LHS}
[\vy-\mM\cdot\vtheta]\T\cdot\tC\inv\cdot[\vy-\mM\cdot\vtheta]
+ [\vtheta-\vmu]\T\cdot\tLambda\inv\cdot[\vtheta-\vmu]
\end{equation}
must equal the quadratic scalar form
\begin{equation}\label{eq:RHS}
[\vtheta-\va]\T\cdot\tA\inv\cdot[\vtheta-\va]
+ [\vy-\vb]\T\cdot\tB\inv\cdot[\vy-\vb]
\hquad,
\end{equation}
because these quadratic scalar forms appear in the exponents in the functions.
This equality follows from straightforward expansion of
all the quadratic forms, plus some use of the matrix inversion lemma\footnote{This
  useful lemma is also called the Woodbury matrix identity. See also \cite{wiki:MIL},
  and \cite{Harville:2011}.}
\begin{equation}\label{eq:invlemma}
[\tQ + \mU\cdot\tS\cdot\mV\T]\inv = \tQ\inv - \tQ\inv\cdot\mU\cdot[\tS\inv + \mV\T\cdot\tQ\inv\cdot\mU]\inv\cdot\mV\T\cdot\tQ\inv
\hquad,
\end{equation}
which gives an expression for the inverse $\tB\inv$ of the marginalized
likelihood variance:
\begin{equation}
\tB\inv = \tC\inv - \tC\inv\cdot\mM\cdot[\tLambda\inv + \mM\T\cdot\tC\inv\cdot\mM]\inv\cdot\mM\T\cdot\tC\inv
\hquad.
\end{equation}
After that it's just a lot of grinding through matrix expressions.\footnote{%
We leave this grinding to the avid reader.
For guidance, it might help to realize that there are terms that
contain $\vtheta\T\cdots\vtheta$, $\vtheta\T\cdots\vy$, $\vy\T\cdots\vy$,
$\vtheta\T\cdots\vmu$, and $\vmu\T\cdots\vmu$.
If you expand out each of these five kinds of terms, each of the five
should lead to an independent-ish equality.}

\paragraph{Solution notes:}
In principle we found this factorization by expanding the quadratic in
(\ref{eq:LHS}) and then completing the square.
Of course we didn't really; we used arguments (which physicists love)
called \emph{detailed balance}:
We required that the terms that look like
$\vtheta\T\cdot\tQ\cdot\vtheta$ were equal between the LHS~(\ref{eq:LHS})
and the RHS~(\ref{eq:RHS}), and then all the terms that look like
$\vmu\T\cdot\tS\cdot\vmu$, and so on.
It turns out you don't have to consider them all to get the right solution.

There is an alternative derivation or proof involving the
\textsl{canonical form} for the multivariate Gaussian. This form is
\begin{equation}
\normal(\vx\given\vm,\tV) = \exp\left(-\frac{1}{2}\vx^T \cdot\tH \cdot \vx + \veta^T \cdot \vx - \frac{1}{2}\,\xi\right)
\end{equation}
\begin{equation}
\tH\equiv\tV\inv \hquad;\hquad
\veta\equiv\tV\inv\cdot\vm \hquad;\hquad
\xi\equiv\ln||2\pi\,\tV||+\veta^T\cdot\tV\cdot\veta \hquad.
\end{equation}
In the canonical form, many products and other manipulations become simpler, so
it is worth trying this route if you get stuck when manipulating Gaussian
expressions.

Because the matrix $\mM$ is not square, it has no inverse. And because this
is a physics problem, $\mM$ has units (which are the units of
$\dd\vy/\dd\vtheta$).
It's beautiful in the solution that $\mM$ and $\mM\T$ appear only where the
units make sense.
They make sense because the units of $\tC\inv$ are inverse data-squared (where $\vy$
is the data vector) and the units of $\tLambda$ are parameters-squared and the units
of $\mM$ are data over parameters.
And they are all different sizes.

If you remember the Bayesian context around \equationname~(\ref{eq:bayes}) and the
Bayesian discussion thereafter,
the Gaussian $\normal(\vtheta\given\va,\tA)$ is the posterior pdf for the linear
parameters $\vtheta$, and the Gaussian $\normal(\vy\given\vb,\tB)$ is the
marginalized likelihood, marginalizing out the linear parameters $\vtheta$.
This marginalization is usually thought of as being an integral, like the
one given in \equationname~(\ref{eq:integral}).
How are these linear-algebra expressions in any sense ``doing this integral''?
The answer is: That integral is a correlation of two Gaussians,\footnote{Astronomers
  like to say that it is the ``convolution'' of two Gaussians, but it is really the
  correlation of two Gaussians. The differences between convolution and correlation
  are minimal, though, and we aren't sticklers.} and the correlation of two
Gaussians delivers a new Gaussian with a shifted mean that is wider than either
of the original two Gaussians.
This factorization does, indeed, deliver the correct marginalization integral.

Continuing along these lines,
various parts of the solution are highly interpretable in terms of the
objects of Bayesian inference. For example, because the term $p(\vtheta\given
\va,\tA)$ is the conditional posterior pdf\footnote{We say ``conditional'' here
  because it is conditioned on nonlinear parameters $\bP$.
  The vector $\va$ and variance tensor
  $\tA$ will depend on the nonlinear parameters $\bP$ through the design matrix
  $\mM$.} for the linear parameters $\vtheta$, the vector $\va$ is the maximum
\foreign{a posteriori} (or \acronym{MAP}) value for the parameter vector
$\vtheta$.
It is found by inverse-variance-weighted combinations of the data and the prior.
In some projects, posterior pdfs or \acronym{MAP} parameter values are the goal
(although we don't think they often should be\footnote{Although---in subjective
  Bayesian inference---the posterior pdf
  is the valid statement of your belief, it is not so useful to your colleagues,
  who start with different beliefs from yours. See \notename\textsuperscript{\ref{note:lf}}.}).
The variance tensor $\tA$ is the posterior variance in the parameter space.
It is strictly smaller (in eigenvalues or determinant) than either
the prior variance $\tLambda$ or the parameter-space data-noise
variance $[\mM\T\cdot\tC\inv\cdot\mM]\inv$.
The vector $\vb$ is the prior-optimal (maximum \foreign{a priori})
value for the data $\vy$.
It is the most probable data vector (prior to seeing any data),
and also the prior expectation for the data,
under the prior pdf.
The variance tensor $\tB$ is the prior variance expanded out to the
data space, and including the noise variance in the data.
It is strictly larger than both the data noise variance $\tC$ and the
data-space prior variance $\mM\cdot\tLambda\cdot\mM\T$.

\paragraph{Implementation notes:}
The solution gives an expression for the variance tensor $\tB$, but
note that when you actually evaluate the pdfs you probably need to
have either the inverse of $\tB$, or else an operator that computes
the product of the inverse and vectors, as in $\tB\inv\cdot\vy$ and
the same for $\vb$.
To get the inverse of the tensor $\tB$ stably, \emph{you might want to use
the matrix inversion lemma} (\ref{eq:invlemma}) given above.
This is often useful because you often know or are given the data inverse variance
tensor $\tC\inv$ for the noise, and the prior variance inverse
$\tLambda\inv$, and the lemma manipulates these into the answer without
any heavy linear algebra.
The lemma saves you the most time and precision when the parameter size $K$
is much smaller than the data size $N$ (or vice versa); that is, when $\mM$
is ``very non-square''.

We also give the general advice that one should \emph{avoid taking an explicit
numerical inverse} (unless you know the inverse exactly in closed form, as you
do for, say, diagonal tensors).
In your code, it is typically stabler to use a \code{solve()} function instead
of the \code{inv()} function.
The reason is that the code operation \code{inv(B)} returns the best
possible inverse to machine precision (if you are lucky), but what you
really want instead is the best possible product of that inverse times
a vector.
So, in general, \code{solve(B,y)} will deliver more precise results than
the mathematically equivalent \code{dot(inv(B),y)}.

The expressions in \equationname~(\ref{eq:problem}) do not require that the variance tensors
$\tC$, $\tLambda$, $\tA$, $\tB$ be positive definite; they only require
that they be non-negative semi-definite.
That means that they can have zero eigenvalues.
As can their inverses $\tC\inv$, $\tLambda\inv$, $\tA\inv$, $\tB\inv$.
If either of these might happen in your problem---like if your prior
freezes the parameters to a subspace of the $\vtheta$-space, which
would lead to a zero eigenvalue in $\tLambda$, or if a data point is
unmeasured or missing, which would lead to a zero eigenvalue in
$\tC\inv$---you might have to \emph{think about how you implement the
  linear algebra operations to be zero-safe}.\footnote{A completely
  zero-safe implementation
  is somewhat challenging, but one comment to make is that if, say, $\tA$
  contains a zero eigenvalue, then there is a direction in the parameter
  space (the $\vtheta$ space) in which the variance vanishes. This means that
  all valid parameter combinations lie on a linear subspace of the full $K$-dimensional
  space. All parameter combinations that wander off the subspace get strictly
  zero probability or negative infinities in the log.
  If your inference is valid, it will probably be the case that the
  vectors at which you want to evaluate \emph{always} lie in the non-zero
  subspace. It makes sense, then, in this case, to work in a representation in which it is easy
  to enforce or ensure that. This usually involves some kind of coordinate transformation
  or rotation or projection. Doing this correctly is beyond the scope of this \documentname.}

\paragraph{Simplification: single multiplicative scaling}

One interesting case is when $K=1$, so the design matrix in fact
reduces to a model vector $\vm$, multiplied by a scalar $\theta$, and
$\mu$ and $\Lambda$ are now scalars as well:
\begin{equation}
\normal(\vy\given\theta\,\vm,\tC)\,\normal(\theta\given\mu,\Lambda) = \normal(\theta\given a,A)\,\normal(\vy\given\vb,\tB)
\hquad.
\end{equation}
This can arise if one wants to multiplicatively scale a model to the data. $a$ would then correspond to the maximum a posteriori value of the multiplicative scaling to fit $\vy$ with $\vm$.
In this case, the previous equations are simplified and no longer
involve many matrix operations. It's a nice exercise to simplify the
solution above for this scalar case.

\paragraph{Special case: wide prior}
Another interesting case that often arises in inferences is the use of an improper
(infinitely wide) prior on the parameters $\vtheta$. Rather than
ignoring the prior pdf on the left-hand side of
\equationname~(\ref{eq:problem}), which is technically incorrect, the correct
posterior pdf can be obtained by taking the limit $\tLambda\inv \rightarrow 0$ in the
fiducial results derived above. It is perhaps
worth noting that in the improper-prior case,
the posterior can still be fine, but the marginalized likelihood will make no sense (it will technically vanish).

\paragraph{Generalization: product of many Gaussians}

A case that arises in some applications is that the
likelihood is made of multiple Gaussian terms, each of which is a
different linear combination of the linear parameters $\vtheta$.
That is, there are $J$ data vectors $\vy_j$, each of which has size or length $N_j$,
and each of which has an expectation
set linearly by the parameters $\vtheta$ but through a different design matrix $\mM_j$.

Provided that the different data vectors $\vy_j$ are independently ``observed'' (that
is, they have independent noise draws with noise variance tensors $\tC_j$),
the total likelihood is just the product of
the individual-data-vector likelihoods.
An example of this case arises in astronomy, for example, when considering
radial velocity measurements of a star taken with different instruments that may
have systematic offsets between their velocity zero-points.

In principle we could work around this problem---reduce it to the previously solved problem---by forming a large vector $\vy$ which is the concatenation of all the individual data vectors $\vy_j$, and a large design matrix $\mM$ which is the concatenation of all the individual design matrices $\mM_j$, and a large total covariance matrix $\tC$ which is a block diagonal matrix containing the noise variance tensors $\tC_j$ on the diagonal blocks.
We could then apply the result of the single-data-vector problem above.
However, this can result in significant unnecessary computation, and it is hard to write the answer in a simple form.
Instead we can take advantage of the separability of the likelihoods, and write the following generalized problem statement:

Find $K$-vector $\va$, $K\times K$ variance tensor $\tA$, $J$ vectors $\vb_j$ (each
of which is a different length $N_j$),
and $J$ variance tensors $\tB_j$ (each of which is a different size $N_j\times N_j$)
such that
\begin{equation}\label{eq:genproblem}
\normal(\vtheta\given\vmu,\tLambda)\,\prod_{j=1}^J\normal(\vy_j\given\mM_j\cdot\vtheta,\tC_j)\,
 = \normal(\vtheta\given\va,\tA)\,\prod_{j=1}^J\normal(\vy_j\given\vb_j,\tB_j) \hquad,
\end{equation}
and such that $\va$, $\tA$, all the $\vb_j$, and all the $\tB_j$
don't depend on $\vtheta$ at all.
Note that
$\vtheta$ is a $K$-vector,
$\vmu$ is a $K$-vector,
$\tLambda$ is a $K\times K$ non-negative semi-definite variance tensor,
each $\vy_j$ is an $N_j$-vector,
each $\mM_j$ is a $N_j\times K$ matrix,
and
each $\tC_j$ is a $N_j\times N_j$ non-negative semi-definite variance tensor.

One way to solve this problem is to write all Gaussians in their canonical form, then separate the elements that depend on $\vtheta$ and on the individual $\vy_j$.
The result can be written as an iteration over data vectors $\vy_j$:
\begin{equation}
  \mbox{\small initialize:}\hquad
  \tA_0\inv = \tLambda\inv \hquad;\hquad \va_0 = \vmu \hquad;\hquad \vx_0 = \tLambda\inv\cdot\vmu
\end{equation}
\begin{align}
  \mbox{\small iterate:}\hquad
  \tB_j &= \tC_j + \mM_j \cdot \tA_{j-1} \cdot \mM_j\T \\
  \vb_j &= \mM_j \cdot \va_{j-1} \\
  \tA_j\inv &= \tA_{j-1}\inv + \mM_j\T \cdot \tC_j\inv \cdot \mM_j \\
  \vx_j &= \vx_{j-1} + \mM_j\T \cdot \tC_j\inv \cdot \vy_j \\
  \va_j &= \tA_j \cdot \vx_j
\end{align}
\begin{align}
  \mbox{\small finish:}\hquad
  \tA &= \tA_J \\
  \va &= \va_J
  \hquad.
\end{align}
The solution is an iteration because you can think of adding each new data
set $\vy_j$ sequentially, with the prior for set $j$ being the posterior from
set $j-1$.
The way this solution is written is unpleasant, because the specific values you get for the
vectors $\vb_j$ and tensors $\tB_j$ depend on the order in which you insert the data.
But---and very importantly for the rules of Bayesian inference---the posterior
mean $\va$ and variance $\tA$ \emph{do not depend} on the order!\footnote{It
  is literally part of the fundamental justification of Bayesian inference
  that the knowledge you eventually have (your final beliefs) does not depend
  on the order in which you observed the data. This is one of the axioms or inputs
  to the theorems that underlie the consistency of Bayesian reasoning.
  There is an illuminating discussion of all this in Chapter~1 of \cite{jaynes}.}

\section{Worked Examples}\label{sec:examples}

When working with a probabilistic model that meets the strong requirements
imposed above (Gaussians everywhere; expectations linear in parameters),
the identities described in this
\documentname\ have practical uses: (1) To simplify the posterior pdf of your
model (which makes generating samples or computing integrals far simpler), and
(2) to reduce the dimensionality of your model (by enabling closed-form
marginalizations over linear parameters).
Reducing the dimensionality of your parameter-space will in general improve
convergence of Markov Chain Monte Carlo\footnote{We have also written a tutorial
on \acronym{MCMC} in this series \citep{Hogg:2018}.} (\acronym{MCMC}) sampling
methods, or enable alternate sampling methods (for example, rejection sampling)
that may be intractable when the parameter dimensionality is large: These two
benefits also typically make inference procedures (like sampling) \emph{far}
faster.
Here, we demonstrate the utility of the identities shown above with two
worked exercises.

\paragraph{Exercise 1: A fully linear model:} We observe
a set of data $(x_i, y_i)$ (indexed by $i$) with
known, Gaussian uncertainties in $y$, $\sigma_y$, and no uncertainty in $x$.
The parametric model we will use for these data is a quadratic polynomial,
\begin{equation}
  f(x \,;\, \alpha, \beta, \gamma) = \alpha\,x^2 + \beta\,x + \gamma
\end{equation}
and we assume we have Gaussian prior pdfs on all of the $K=3$ linear parameters
$(\alpha, \beta, \gamma)$,
\begin{align}
  p(\alpha) &= \normal(\alpha \given \mu_\alpha, \sigma_\alpha)\\
  p(\beta) &= \normal(\beta \given \mu_\beta, \sigma_\beta)\\
  p(\gamma) &= \normal(\gamma \given \mu_\gamma, \sigma_\gamma)
  \hquad.
\end{align}
While this example may seem overly simple or contrived, quadratic models are
occasionally useful in astronomy and physics, for example, when centroiding a
peak,\footnote{Fitting second-order polynomials has been shown to be great for
  centroiding peaks in astronomy contexts. See, for example, \cite{vakili}, and
  \cite{teague}.}
and polynomial models are often used to capture smooth trends in
data.

\marginpar{\tfigurerule\\
  \parbox{\marginparwidth}{
    ~\hfill\begin{tabular}{c|c|c}
      $x$ & $y$ & $\sigma_y$ \\
      \hline
      $-0.6$ & 12.2 & 0.8 \\
      2.0 & 4.1 & 3.2 \\
      2.7 & 0.9 & 3.3 \\
      3.6 & $-15.0$ & 3.9 \\
    \end{tabular}\hfill~}
  \includegraphics[width=\marginparwidth]{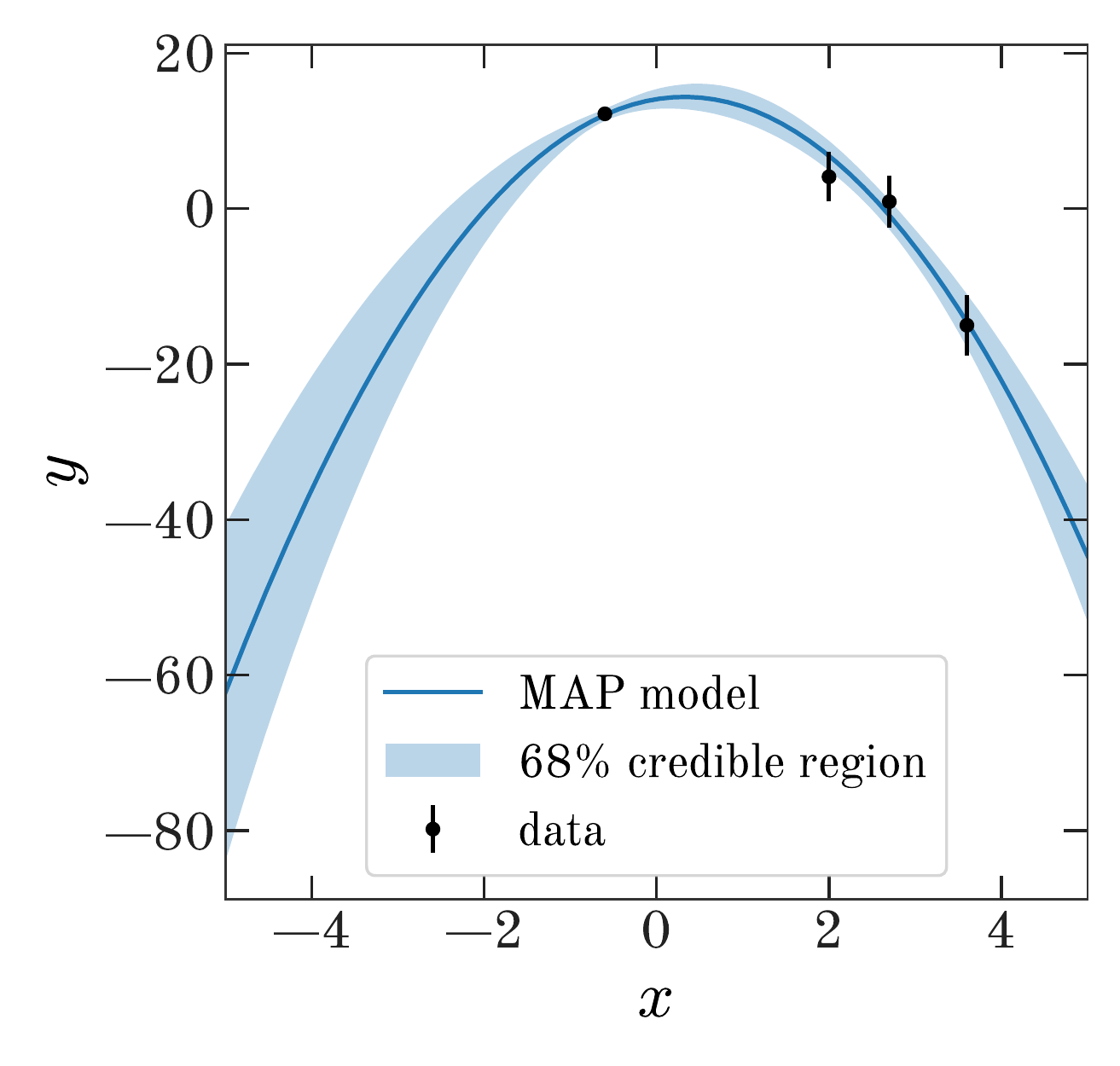}
  \caption{\textsl{Top:} Data generated with ``true'' parameters
    $(\alpha, \beta, \gamma) = (3.21, 2.44, 14.82)$.
    \textsl{Bottom:} The solution to Exercise 1. The data points (black markers)
    show the data from the table. The \acronym{MAP} parameter values were
    found using \equationname~(\ref{eq:a}).
    All data files and solution notebooks are available via
    \href{https://doi.org/10.5281/zenodo.3855689}{Zenodo} \citep{zenodo}.\label{fig:ex1}}\\
  \bfigurerule}
The data table shown in \figurename~\ref{fig:ex1} contains $N=4$ data points,
$(x_i, y_i, \sigma_{y_i})$, generated using this quadratic model.
Assuming values for the prior means, $\vmu$, and prior variance tensor,
$\tLambda$,
\begin{align}
  \vmu\T &= (\mu_\alpha, \mu_\beta, \mu_\gamma) = (1, 3, 9)\\
  \tLambda &=
    \begin{pmatrix}
      5^2 & 0 & 0\\
      0 & 2^2 & 0\\
      0 & 0 & 8^2
    \end{pmatrix}
\end{align}
compute the \acronym{MAP} parameter values $\va\T = (\alpha_{\rm MAP}, \beta_{\rm MAP},
\gamma_{\rm MAP})$.
Plot the data (with error bars) and over-plot the model evaluated at the
\acronym{MAP} parameter values.
Generate 4096 posterior samples of the linear parameters.
Over-plot a shaded region showing the 68~percent credible region for the model,
estimated using these samples.

\paragraph{Solution:} Given the assumptions and prior parameter values
above, the design matrix, $\mM$, is
\begin{align}
  \mM &= \begin{pmatrix}
      0.36 & -0.6 & 1\\
      4.0 & 2.0 & 1\\
      7.29 & 2.7 & 1\\
      12.96 & 3.6 & 1
    \end{pmatrix} \hquad.
\end{align}
By plugging in to \equationname~(\ref{eq:a}), we find \acronym{MAP} parameter values for the
linear parameters
\begin{equation}
  \va\T =
    (\alpha_{\rm MAP}, \beta_{\rm MAP}, \gamma_{\rm MAP}) =
      (3.61, 1.98, 14.26) \hquad.
\end{equation}
\figurename~\ref{fig:ex1} shows the data (black points), the model computed with
the \acronym{MAP} parameter values (blue line), and the 68-percent credible region (shaded
blue region) estimated using posterior samples generated from
$\normal(\vtheta\given\va,\tA)$.
The companion \href{https://doi.org/10.5281/zenodo.3855689}{\texttt{IPython}
notebook} (\texttt{Exercise1.ipynb}) contains the full solution.

\paragraph{Exercise 2: A model with a nonlinear parameter:} We
observe a set of data $(x_i, y_i)$ (indexed by $i$) with
known, Gaussian uncertainties in $y$, $\sigma_{y_i}$, and no uncertainty in $x$.
The parametric model we will use for these data is a generalized sinusoid with a
constant offset,
\begin{equation}
  f(x \,;\, \alpha, \beta, \gamma, \omega) =
    \alpha\,\cos(\omega \, x) + \beta\,\sin(\omega \, x) + \gamma \label{eq:ex2model}
\end{equation}
and we again assume we have Gaussian prior pdfs on all of the \emph{linear}
parameters $(\alpha, \beta, \gamma)$.
Models like this (a periodic model with both linear and nonlinear parameters)
are common in astronomy, especially in the context of asteroseismology, light
curve analysis, and radial velocity variations from massive companions (binary
star systems or exoplanets).
For this setup, we can no longer analytically express the posterior pdf because
of the nonlinear parameter $\omega$, but we can compute the marginal likelihood
(marginalizing over the linear parameters) conditioned on the frequency
$\omega$.

\marginpar{\tfigurerule\\
  \parbox{\marginparwidth}{
    ~\hfill\begin{tabular}{c|c|c}
    $x$ & $y$ & $\sigma_y$ \\
    \hline
    $-1.2$ & 11.2 & 0.2 \\
    1.3 & 16.1 & 0.2 \\
    3.1 & 10.2 & 0.3 \\
    4.1 & 13.5 & 0.3
    \end{tabular}\hfill~}
  \includegraphics[width=\marginparwidth]{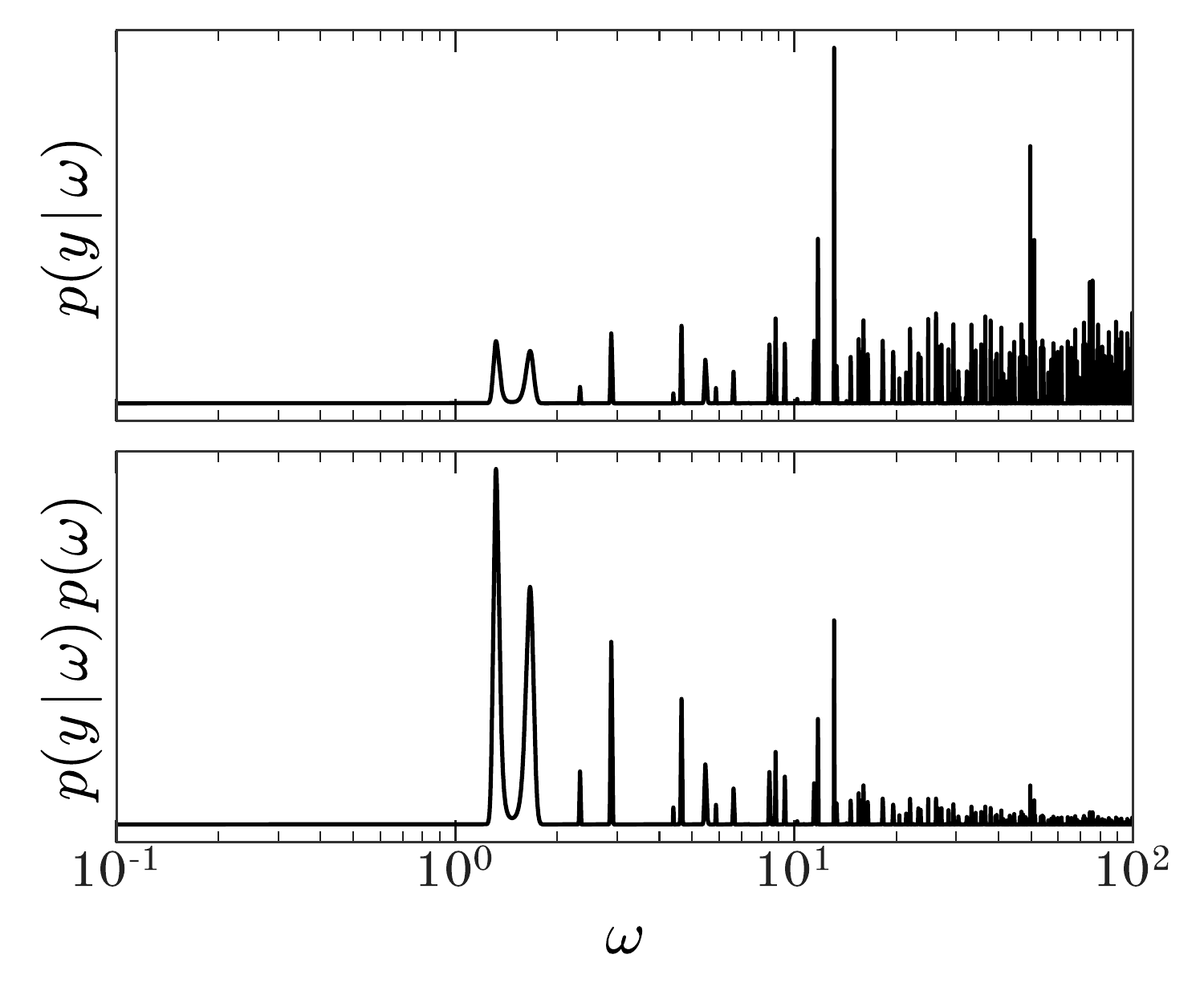}
  \includegraphics[width=\marginparwidth]{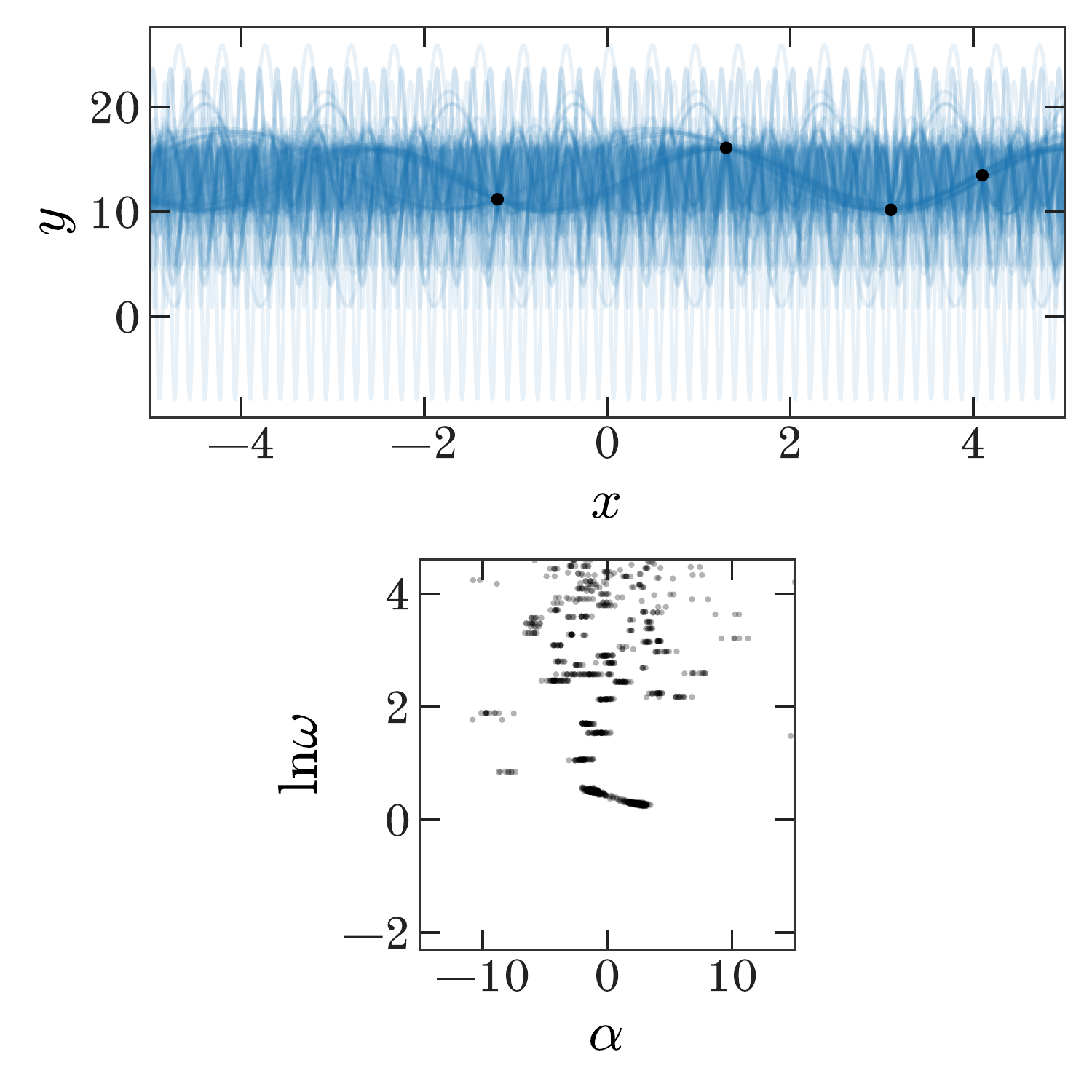}
  \caption{\textsl{Top:} Data generated with ``true'' parameters
    $(\alpha, \beta, \gamma, \omega) = (3.21, 2.44, 13.6, 1.27)$.
    \textsl{Middle and Bottom:} The solution to Exercise 2.
    All data files and solution notebooks are available via
    \href{https://doi.org/10.5281/zenodo.3855689}{Zenodo} \citep{zenodo}.\label{fig:ex2}}\\
  \bfigurerule}
The table shown in \figurename~\ref{fig:ex2} contains $N=4$ data points,
$(x_i, y_i, \sigma_{y_i})$, generated with this sinusoid model.
Assuming values for the prior means, $\vmu$, and prior variance tensor,
$\tLambda$,
\begin{align}
  \vmu\T &= (\mu_\alpha, \mu_\beta, \mu_\gamma) = (0, 0, 0)\\
  \tLambda &=
    \begin{pmatrix}
      5^2 & 0 & 0\\
      0 & 5^2 & 0\\
      0 & 0 & 10^2
    \end{pmatrix}
\end{align}
write a function to compute the vectors and matrices we need for the linear
parameters (the design matrix and components
$\va, \tA, \vb, \tB$
of the factorization) at a given value of the frequency $\omega$.
Assuming a prior on $\omega$ that is uniform in $\ln\omega$ over the domain
$(0.1, 100)$,
\begin{equation}
  p(\omega) \propto \frac{1}{\omega}
\end{equation}
evaluate the log-marginal likelihood $\ln p(\vy\given\omega)$ and add to the
log-frequency prior $\ln p(\omega)$ over a grid of 16,384 frequencies $\omega$
between (0.1, 100).
Plot both the marginal likelihood (not log!) $p(\vy\given\omega)$ and the
posterior pdf $p(\vy\given\omega)\,p(\omega)$ as a function of this frequency
grid.

Generate 512 posterior samples\footnote{The
posterior pdf over $\omega$ will be extremely multimodal. Don't fire up
standard \acronym{MCMC}! Try using rejection sampling instead: Generate a
dense prior sampling in the nonlinear parameter $\omega$, evaluate
the marginalized likelihood at each sample in $\omega$, and use this
to reject prior samples.} in the full set of parameters
$(\alpha,\beta,\gamma,\omega)$.
Make a scatter plot showing a 2D projection of these samples in
$(\alpha, \ln \omega)$.
Plot the data, and over-plot 64 models (\equationname~\ref{eq:ex2model}) computed
using a fair subset of these posterior samples.
The companion \href{https://doi.org/10.5281/zenodo.3855689}{\texttt{IPython}
notebook} (\texttt{Exercise2.ipynb}) contains the full solution.

\clearpage\raggedright
\bibliography{refs}

\end{document}